\documentclass[preprint2]{aastex6}
\usepackage{xspace}
\usepackage{amsmath}

\usepackage[T1]{fontenc}
\usepackage{ifthen}

\newcommand{\Kepler}{\textit{Kepler}\xspace}

\newcommand{\Gaia}{\textit{Gaia}\xspace} 
\newcommand{\Mstar}{\ensuremath{M_{\star}}\xspace}
\newcommand{\Rstar}{\ensuremath{R_{\star}}\xspace} 
 
\newcommand{\fe}{\ensuremath{\mathrm{[Fe/H]}}\xspace}
\newcommand{\teff}{\ensuremath{T_{\mathrm{eff}}}\xspace}  
 
\newcommand{\vsini}{\ensuremath{v \sin i}\xspace}

\newcommand{\Mp}{\ensuremath{M_{P}}\xspace}

\newcommand{\Rp}{\ensuremath{R_P}\xspace}

\newcommand{\ms}{\ensuremath{\mathrm{m}\,\mathrm{s}^{-1}}\xspace}
\newcommand{\kms}{\ensuremath{\mathrm{km}\,\mathrm{s}^{-1}}\xspace}

\newcommand{\AU}{\ensuremath{\mathrm{AU}}\xspace}

\newcommand{\Me}{\ensuremath{M_{\oplus}}\xspace} 
\renewcommand{\Re}{\ensuremath{R_{\oplus}}\xspace}

\newcommand{\Rsun}{\ensuremath{R_{\odot}}\xspace}
\newcommand{\Msun}{\ensuremath{M_{\odot}}\xspace}

\newcommand{\bjdtdb}{\ensuremath{\mathrm{BJD}_\mathrm{TBD}}\xspace}
\def\deg{\ensuremath{^{\circ}}}
 
\newcommand{\angstrom}{\AA\xspace}
\newcommand{\uas}{\ensuremath{\mu \mathrm{as}}\xspace}

\newcommand{\sigjit}[1]{
        \ensuremath{
                \ifthenelse{\equal{#1}{}}{\sigma_\mathrm{jit}}{\sigma_\mathrm{jit,#1}}}
        \xspace
}

\newcommand{\loglike}{\ensuremath{\ln \mathcal{L}}\xspace}

\newcommand{\dbic}{\ensuremath{\Delta\mathrm{BIC}}\xspace}
\usepackage{xstring} % Needed for IfEqCase

\newcommand{\nobsraw}{253\xspace}
\newcommand{\nobs}{144\xspace}

\newcommand{\Shk}{\ensuremath{\mathrm{S}_{\mathrm{HK}}}\xspace}
\newcommand{\Halpha}{\ensuremath{\mathrm{H}\alpha}\xspace}

\newcommand{\singleplanetmodel}[1]{%
  \IfEqCase{#1}{%
  {P1}{$\equiv$ 4.887802443}
  {T01}{$\equiv$ 2454957.8132067}
  {K1}{10.0$^{+2.2}_{-2.3}$}
  {e1}{0.19$^{+0.14}_{-0.13}$}
  {w1}{33$^{+44}_{-57}$}
  {c1}{92$^{+43}_{-44}$}
  {gamma}{-10.2$\pm 1.5$}
  {dvdt}{-0.0071$\pm 0.0012$}
  {jit}{17.3$^{+1.1}_{-1.0}$}
  {rms}{16.86}
  {bic}{1250.91}
  }[\PackageError{tree}{Undefined option to tree: #1}{}]%%
}%

\newcommand{\twoplanetmodel}[1]{%
  \IfEqCase{#1}{%
  {P1}{$\equiv$ 4.887802443}
  {T01}{$\equiv$ 2454957.8132067}
  {K1}{10.51$^{+0.72}_{-0.71}$}
  {e1}{0.223$^{+0.036}_{-0.034}$}
  {w1}{18$^{+15}_{-17}$}
  {P2}{3335$^{+280}_{-160}$}
  {T02}{2456739$^{+30}_{-39}$}
  {K2}{32.1$\pm 1.4$}
  {e2}{0.560$^{+0.034}_{-0.035}$}
  {w2}{140.0$^{+4.8}_{-4.9}$}
  {c1}{$\equiv$ 0}
  {gamma}{-1.01$^{+0.97}_{-0.94}$}
  {dvdt}{-0.0011$^{+0.002}_{-0.0014}$}
  {jit}{5.40$^{+0.38}_{-0.34}$}
  {rms}{5.37}
  {bic}{941.99}
  }[\PackageError{tree}{Undefined option to tree: #1}{}]%%
}%

\newcommand{\twoplanetdecorr}[1]{%
  \IfEqCase{#1}{%
  {P1}{$\equiv$ 4.887802443}
  {T01}{$\equiv$ 2454957.8132067}
  {K1}{10.42$^{+0.64}_{-0.66}$}
  {e1}{0.218$^{+0.034}_{-0.031}$}
  {w1}{19$^{+14}_{-16}$}
  {M1}{$23.4 \pm 1.5$}
  {a1}{0.05254$^{+0.00064}_{-0.00066}$}
  {rp1}{0.0413$^{+0.0018}_{-0.0019}$}
  {tp1}{2454957.15$^{+0.17}_{-0.20}$}
  {ra1}{0.0637$^{+0.0020}_{-0.0019}$}
  {ta1}{2454959.60$^{+0.17}_{-0.20}$}
  {P2}{3407$^{+360}_{-190}$}
  {P2yr}{9.33$^{+0.99}_{-0.52}$}
  {T02}{2456746$^{+24}_{-32}$}
  {K2}{30.9$\pm 1.3$}
  {e2}{0.601$^{+0.032}_{-0.031}$}
  {w2}{143.7$^{+4.8}_{-4.9}$}
  {M2}{$507^{+30}_{-27}$}
  {M2MJ}{$1.60^{+0.09}_{-0.08}$}
  {a2}{4.13$^{+0.29}_{-0.16}$}
  {rp2}{1.67$^{+0.14}_{-0.13}$}
  {tp2}{2456862$^{+20}_{-26}$}
  {ra2}{6.61$^{+0.52}_{-0.30}$}
  {ta2}{2458565$^{+166}_{-87}$}
  {sep2}{5.09$^{+0.55}_{-0.38}$}
  {angsep2}{134$^{+15}_{-10}$}
  {astrom}{264$^{+42}_{-29}$}
  {c1}{78.6$^{+16.8}_{-16.1}$}
  {gamma}{-1.80$^{+0.84}_{-0.79}$}
  {dvdt}{0.0001$^{+0.0018}_{-0.0013}$}
  {jit}{4.98$^{+0.36}_{-0.32}$}
  {rms}{4.98}
  {bic}{925.59}
  }[\PackageError{tree}{Undefined option to tree: #1}{}]%%
}%

\newcommand{\twoplanetpartial}[1]{%
  \IfEqCase{#1}{%
  {P1}{$\equiv$ 4.887802443}
  {T01}{$\equiv$ 2454957.8132067}
  {K1}{10.87$^{+0.67}_{-0.66}$}
  {e1}{0.223$^{+0.035}_{-0.032}$}
  {w1}{22$^{+13}_{-15}$}
  {P2}{4059$^{+820}_{-480}$}
  {T02}{2456712$^{+24}_{-33}$}
  {K2}{33.2$\pm 1.5$}
  {e2}{0.584$^{+0.04}_{-0.033}$}
  {w2}{137.1$^{+4.4}_{-4.1}$}
  {c1}{108.3$\pm 21.6$}
  {gamma}{-1.94$^{+0.71}_{-0.65}$}
  {dvdt}{0.0034$^{+0.0027}_{-0.0025}$}
  {jit}{4.80$^{+0.36}_{-0.31}$}
  {rms}{4.82}
  {bic}{853.32}
  }[\PackageError{tree}{Undefined option to tree: #1}{}]%%
}%

\newcommand{\stellar}[1]{%
  \IfEqCase{#1}{%
  {mass}{$0.809^{+0.02}_{-0.03}$}
  {radius}{$0.683 \pm 0.009$}
  {distance}{$37.89 \pm 0.33$}
  {teff}{$4780 \pm 50$}
  {fe}{$+0.31 \pm 0.05$}
  {Vmag}{$6.57 \pm 0.09$}
  {vsini}{$1.5 \pm 1.5$}
  {prot}{29.2}
  {age}{$6.5^{+5.9}_{-4.1}$}
  }[\PackageError{tree}{Undefined option to tree: #1}{}]%%
}%

\newcommand{\val}[2]{%
  \IfEqCase{#1}{%
    {single-planet-model}{\singleplanetmodel{#2}}%
    {two-planet-model}{\twoplanetmodel{#2}}%
    {two-planet-decorr}{\twoplanetdecorr{#2}}%
    {two-planet-partial}{\twoplanetpartial{#2}}%
    {stellar}{\stellar{#2}}%
  }[\PackageError{tree}{Undefined option to tree: #1}{}]%
}%

\begin{document}


\title{HAT-P-11: Discovery of a Second Planet and a Clue to Understanding Exoplanet Obliquities}

%% Use \author, \affil, plus the \and command to format author and affiliation 
%% information.  If done correctly the peer review system will be able to
%% automatically put the author and affiliation information from the manuscript
%% and save the corresponding author the trouble of entering it by hand.
%%
%% The \affil should be used to document primary affiliations and the
%% \altaffil should be used for secondary affiliations, titles, or email.

%% Authors with the same affiliation can be grouped in a single

\author{Samuel W.\ Yee\altaffilmark{1,12}}
\author{Erik A.\ Petigura\altaffilmark{1,7}}
\author{Benjamin J.\ Fulton\altaffilmark{1,8}}
\author{Heather A.\ Knutson\altaffilmark{1}}
\author{Konstantin Batygin\altaffilmark{1}}
% Alphabetical after here
\author{G\'asp\'ar \'A. Bakos\altaffilmark{2,9}}
\author{Joel D.\ Hartman\altaffilmark{2}}
\author{Lea A.\ Hirsch\altaffilmark{3}}
\author{Andrew W. Howard\altaffilmark{1}}
\author{Howard Isaacson\altaffilmark{3}}
\author{Molly R.\ Kosiarek\altaffilmark{4,10}}
\author{Evan Sinukoff\altaffilmark{5,1}}
\and
\author{Lauren M.\ Weiss\altaffilmark{6,11}}

\altaffiltext{1}{California Institute of Technology, Pasadena, CA, 91125, USA}
\altaffiltext{2}{Department of Astrophysical Sciences, Princeton University, Princeton, NJ 08544, USA}

\altaffiltext{3}{University of California, Berkeley, Berkeley, CA 94720, USA}
\altaffiltext{4}{University of California, Santa Cruz, Santa Cruz, CA, 95064, USA}
\altaffiltext{5}{Institute for Astronomy, University of Hawai`i, Honolulu, HI 96822, USA} 
\altaffiltext{6}{University of Montr\'{e}al, Montr\'{e}al, QC H3T 1J4, Canada}
\altaffiltext{7}{Hubble Fellow}
\altaffiltext{8}{Texaco Postdoctoral Fellow}
\altaffiltext{9}{Packard Fellow}
\altaffiltext{10}{NSF Graduate Research Fellow}
\altaffiltext{11}{Trottier Fellow}
\altaffiltext{12}{syee@caltech.edu}

\begin{abstract}
HAT-P-11 is a mid-K dwarf that hosts one of the first Neptune-sized planets found outside the solar system. The orbit of HAT-P-11b is misaligned with the star's spin --- one of the few known cases of a misaligned planet orbiting a star less massive than the Sun. We find an additional planet in the system based on a decade of precision radial velocity (RV) measurements from Keck/HIRES. HAT-P-11c is similar to Jupiter in its mass ($\Mp\sin{i} = 1.6\pm0.1$~$M_J$) and orbital period ($P = 9.3^{+1.0}_{-0.5}$~year), but has a much more eccentric orbit ($e=0.60\pm0.03$). In our joint modeling of RV and stellar activity, we found an activity-induced RV signal of $\sim$7~\ms, consistent with other active K dwarfs, but significantly smaller than the 31~\ms reflex motion due to HAT-P-11c. We investigated the dynamical coupling between HAT-P-11b and c as a possible explanation for HAT-P-11b's misaligned orbit, finding that planet-planet Kozai interactions cannot tilt planet b's orbit due to general relativistic precession; however, nodal precession operating on million year timescales is a viable mechanism to explain HAT-P-11b's high obliquity. This leaves open the question of why HAT-P-11c may have such a tilted orbit. At a distance of 38~pc, the HAT-P-11 system offers rich opportunities for further exoplanet characterization through astrometry and direct imaging.
\end{abstract}


\section{Introduction}
\label{sec:intro}
HAT-P-11 is a mid-K dwarf known to host HAT-P-11b, a super-Neptune on a $P=4.88$~day orbit, with $\Mp = $~\val{two-planet-decorr}{M1}~\Me and $\Rp=4.36 \pm 0.06$~\Re. The planet was first discovered by \cite{Bakos10} using ground-based photometry and confirmed by radial velocities (RVs), which constrained its mass and eccentricity. \cite{Bakos10} found a moderate eccentricity of $e = 0.198 \pm 0.046$, the first clue that the HAT-P-11 system is dynamically hot. At the time, HAT-P-11b was the smallest planet discovered by ground-based transit photometry.

HAT-P-11 was observed by the {\em Kepler Space Telescope} \citep{Borucki10} during its prime mission (2009--2013). \cite{Deming11} and \cite{SanchisOjeda11} analyzed this data and found spot-crossing anomalies at particular phases of the transit of HAT-P-11b, which are consistent with a nearly polar orbit crossing two active latitudes on the host star. This was in agreement with the results from two independent RV campaigns by \cite{Winn10} and \cite{Hirano11}, who used the Rossiter-McLaughlin (RM) effect to measure the planet's orbital obliquity to be $\lambda\approx100\deg$. Using the \Kepler photometry, \cite{Huber17} also reported a tentative detection of HAT-P-11b's secondary eclipse.

Here, we present an extended RV timeseries spanning 10 years (Section \ref{sec:obs}), which show a long-period Keplerian signal with $P \approx 9$ years. While HAT-P-11 is chromospherically active, we show in Section \ref{sec:planet_evidence} that the RV signal cannot be explained by activity alone. In Section \ref{sec:joint_modeling}, we model the RV time series including the effects of planet b, planet c, and stellar activity. We investigate the dynamical connection between the two planets in Section \ref{sec:dynamics} and find that HAT-P-11c can explain the high obliquity of HAT-P-11b. Finally, we place the HAT-P-11 system in context of other exoplanet systems (Section \ref{sec:context}) and discuss prospects for future characterization (Section \ref{sec:future_obs}).

\section{Spectroscopic Observations}
\label{sec:obs}

The California Planet Search (CPS; \citealt{Howard10}) has observed the HAT-P-11 system since 2007 August with the High Resolution Echelle Spectrometer (HIRES; \citealt{Vogt94}) at the Keck I 10m telescope on Maunakea. We collected a total of \nobsraw spectra with an iodine cell in front of the spectrometer, which imprints iodine absorption lines to serve as a wavelength reference against which RVs can be measured precisely. The spectra have signal-to-noise ratios (S/N) between 100 and 130 per pixel on blaze near 5500~\angstrom.

\subsection{Radial Velocities}
\label{ssec:rvs}

We used the standard CPS pipeline described in \cite{Howard10} to determine the RVs. This involves forward modeling the stellar and iodine spectra convolved with the instrumental point spread function for different spectral segments \citep{Marcy92,Valenti95}. The complete set of RV data is presented in Table \ref{tab:rv}, with a median uncertainty of $1.4$~\ms. In the subsequent analysis, we have excluded two sets of very high cadence observations taken within 4~hr of the transit of HAT-P-11b, which are affected by the RM effect.

\begin{figure*}
	\epsscale{1.2}
	\plotone{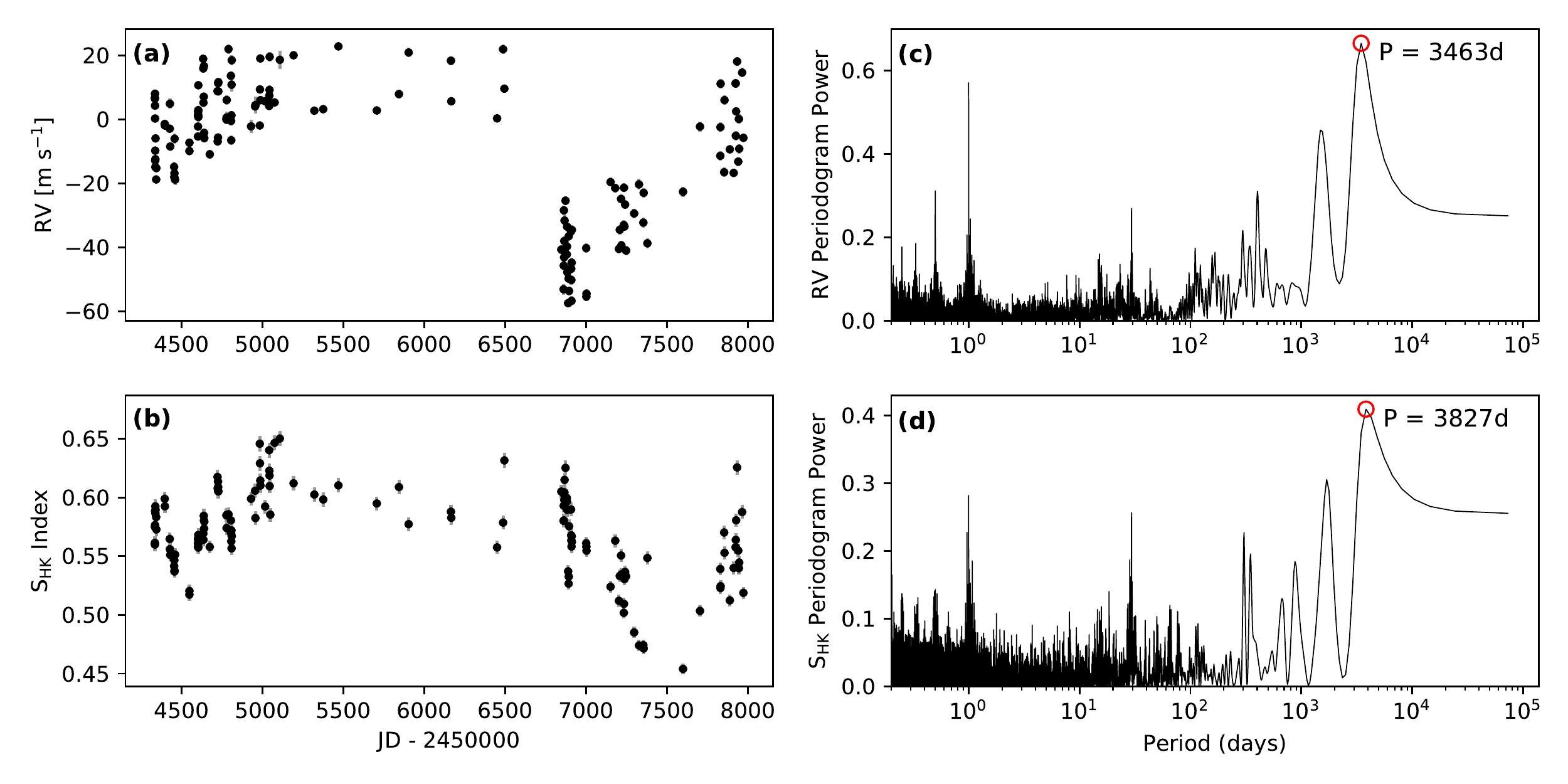}
	\caption{Panel (a): Radial velocity (RV) time series, showing a long-period signal.
	Panel (b): \Shk index time series. Error bars for both panels (a) and (b) are shown but are comparable to the size of the points.
	Panel (c): A Lomb-Scargle periodogram of the RV data shows strong power at $P\approx3500$~days. This signal and its harmonics dominate the periodogram and overwhelm the 4.88~day signal from the known inner planet, HAT-P-11b.
	Panel (d): The \Shk periodogram has a peak at a similar period. We note the strong signal in both periodograms at 29 days, the inferred rotational period of HAT-P-11.
	\label{fig:rv_timeseries}}
\end{figure*}

This leaves us with \nobs remaining RV measurements, which are plotted in Figure \ref{fig:rv_timeseries}a. In their original discovery, \cite{Bakos10} reported a significant long-term drift over two years of RV observations, which they interpreted as a possible second planet. With our extended observational baseline of ten years, we see that this long-period trend has reversed, suggesting that we have now viewed a complete orbit of this outer companion. A generalized Lomb-Scargle periodogram \citep{Zechmeister09} of the raw RVs shows a peak at $\sim3463$~days (Figure \ref{fig:rv_timeseries}c), just over 9~years.

\begin{deluxetable*}{lRRRRRRR}
\tablecaption{Radial Velocity and Activity Measurements \label{tab:rv}}
\tablecolumns{7}
\tablewidth{0pt}
\tablehead{
	\colhead{Time} &
	\colhead{RV} &
	\colhead{$\sigma$(RV)} &
	\colhead{\Shk Index} &
	\colhead{$\sigma$(\Shk)} &
	\colhead{\Halpha Index} &
	\colhead{$\sigma$(\Halpha)} &
	\colhead{Flag} \\
	\colhead{\bjdtdb} &
	\colhead{\ms} &
	\colhead{\ms} &
	\colhead{} &
	\colhead{} &
	\colhead{} &
	\colhead{} &
	\colhead{}
}
\startdata
% time, mnvel, sig_vel, sval, sig_sval, ha, sig_ha, mask
2454335.891030 & 6.50 & 1.03 & 0.5599 & 0.0056 & 0.04539 & 0.00026 &1 \\
2454335.897680 & 6.75 & 1.09 & 0.5614 & 0.0056 & 0.04537 & 0.00026 &1 \\
2454336.746470 & 8.03 & 0.94 & 0.5748 & 0.0057 & 0.04533 & 0.00025 &1 \\
2454336.859340 & 4.30 & 1.03 & 0.5751 & 0.0058 & 0.04531 & 0.00026 &1 \\
2454336.947330 & 0.27 & 1.00 & 0.5765 & 0.0058 & 0.04543 & 0.00027 &1 \\
2454337.729220 & -12.86 & 1.14 & 0.5886 & 0.0059 & 0.04602 & 0.00028 &1 \\

\enddata
\tablecomments{Radial velocity (RV) and activity measurements calculated from HIRES observations. A 1 in the Flag column indicates that the data point was used in our analysis. Table \ref{tab:rv} is published in its entirety in machine-readable format. A portion is shown here for guidance regarding its form and content.}
\end{deluxetable*}

\subsection{Stellar Activity Indicators}
\label{ssec:activity_indicators}

HAT-P-11 is known to be a spotted, chromospherically active star \citep{Deming11,Morris17a}. Stellar activity can produce spurious RV signals that may be mistaken for a planet (see e.g. \citealt{Robertson14a,Haywood14}). To investigate whether stellar activity could account for the 9~year RV signal, we extract two activity indices from our spectroscopic observations.

The Mount Wilson \Shk index traces the chromospheric emission in the cores of the Ca II H\&K lines \citep{Vaughan78} and is a standard activity tracer for main-sequence stars. We extract \Shk from our spectra following the procedure of \cite{Isaacson10}, and our measurements are precise to 1\%. 

We also measured the \Halpha index, which has been found to be a good activity tracer for late-type stars \citep{DaSilva11,Robertson14a}. While \Halpha tracked \Shk closely, the size of the variations were on the 1-2\% level, comparable to the measurement uncertainty. Therefore, we henceforth use \Shk as the activity tracer. Both activity indices for each observation are provided in Table \ref{tab:rv}. 

The periodogram of \Shk (Figure \ref{fig:rv_timeseries}d) has a peak at $\sim3800$~days, close to peak found in the RV periodogram. \cite{Morris17b, Morris17a} also observed this activity signal and interpreted it as a solar-like dynamo. Given the comparable timescales of the RV and activity cycles, we consider whether activity could be responsible for the RV variability in the following section.

\section{Is the Long-Period RV Signal Due to Stellar Activity?}
\label{sec:planet_evidence}

Our decade of RV observations of HAT-P-11 have revealed a long-period signal, suggestive of a planet. Here, we assess whether this signal could be caused by stellar activity. We show that activity is incompatible with the observed 9 year RV signal for three reasons: (1) the observed amplitude is much larger than activity-induced RV variability seen in similar stars, (2) there is a significant phase offset between the activity and RV cycles, and (3) the RV-activity correlation is too weak to account for the RV signal. 

\subsection{Amplitude of RV Signal}
\label{ssec:rvmag}

We first subtracted the effect of HAT-P-11b from the RV time series, using a model generated from the orbital parameters derived by \cite{Bakos10}. The residual RVs are shown in Figure \ref{fig:keplerian_sval} and the remaining long-period signal has a semi-amplitude of $\sim35$~\ms.

Typical activity-induced RV signals are significantly smaller. \cite{Isaacson10} measured the chromospheric activity and RV jitter of $\sim2600$ main-sequence and subgiant stars. For the $\sim300$~stars in the sample similar to HAT-P-11 ($1.0 < B-V < 1.3$), the typical RMS in measured RVs was around $4-8$~\ms. In particular, there was no increase in jitter as a function of \Shk index, suggesting that K dwarfs do not have significant activity-induced jitter.

These findings were corroborated in a similar study by \cite{Lovis11}, who observed 304 FGK stars with HARPS over seven years, finding a maximum activity-induced RV signal of 11~\ms. This study also found that RV correlation with magnetic activity is minimized in stars with $\teff~\approx4800$~K, where even strong magnetic cycles induced RV signals of only several \ms. 

Thus, it is unlikely that the $\sim35$~\ms RV signal in the HAT-P-11 data could be attributed to stellar activity alone, as it is more than three times larger than previously known activity-induced signals, particularly when we consider the reduced sensitivity of RVs to chromospheric activity in K dwarfs.

\subsection{RV-Activity Phase Offset}
\label{ssec:phase_mismatch}

Another line of reasoning favoring the planet interpretation is the phase offset between the RV and \Shk cycles. Activity-induced RV signals arise due to suppression of convective blueshift, primarily by plages (e.g. \citealt{Haywood14,Dumusque14}). Because the \Shk index measures chromospheric Ca II H\&K emission, it is a direct measure of plage activity \citep{Shine74}. Hence, any activity-induced RV signal should move in lockstep with the \Shk activity indicator, without any phase offset. 

The presence of spots may cause a phase shift between the \Shk and induced RV signal, due to masking of parts of the star that are rotationally blue- or red-shifted \citep{Haywood14}. The maximum offset between the two signals due to rotation is only a fraction of the stellar rotation period (29~days) and is therefore negligible when compared to the 9~year period of the RV signal.

\begin{figure}
	\epsscale{1.2}
	\plotone{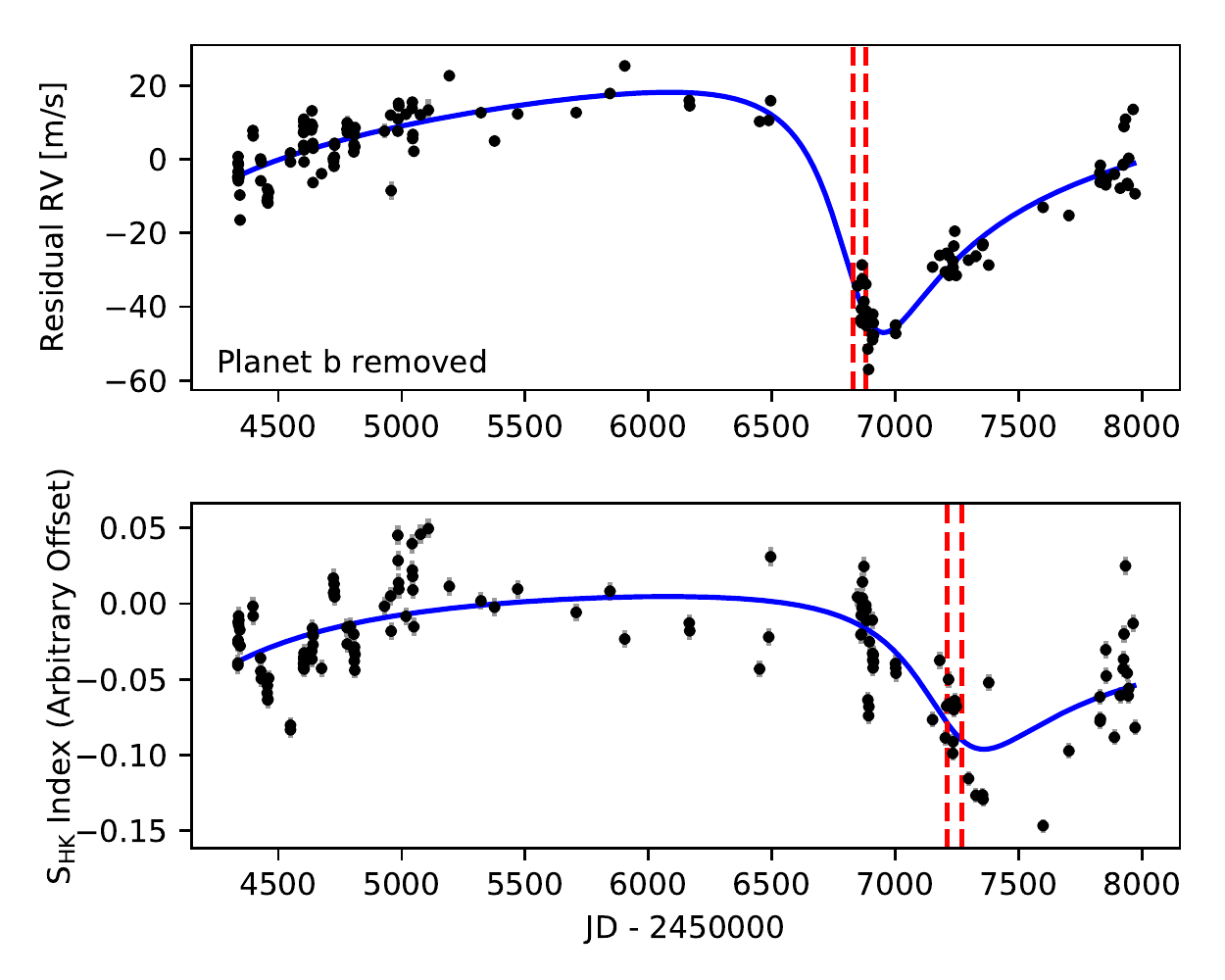}
	\caption{
	Top: The maximum-likelihood Keplerian model fit to the residual RVs, after removing the effect of the inner planet. Vertical dashed lines mark the 1-$\sigma$ confidence interval for the time of periastron passage.
	Bottom: Using the model parameters and uncertainties derived from the RV fit as priors, we fit a Keplerian to the \Shk indices. The time of periastron passage for this model is 412 days later, demonstrating a significant phase offset between the two signals. \label{fig:keplerian_sval}}
\end{figure}

Inspection of Figure \ref{fig:rv_timeseries} shows that the RV and \Shk time series reach their respective minima at times that differ by more than a year. To measure the significance of this offset, we used the publicly available \texttt{RadVel} software package \citep{Fulton18}\footnote{\url{https://radvel.readthedocs.io}} to fit a Keplerian model to the residual RVs described in Section \ref{ssec:rvmag}, and to the \Shk indices (Figure \ref{fig:keplerian_sval}).

For the RVs, we measured an eccentricity of $e=0.565\pm0.035$, period of $3334\pm220$~days, and a periastron passage of JD~$=2456859^{+22}_{-31}$. If this signal were in fact due to stellar activity, we would expect the shape and period of the \Shk cycle to be similar. We thus fit the \Shk time series with another Keplerian using priors on eccentricity and period corresponding to the RV fit. For the \Shk indices, we measure a ``periastron passage'' of JD~$=2457271^{+28}_{-34}$, more than 400 days after $t_p$ of the RV signal. This corresponds to a phase offset of $\sim12\%$, a difference of $>10\sigma$.

There is no physical basis to expect such a 400~day offset between the long-period \Shk and RV cycles. This suggests that their apparent similarity is no more than a coincidence, rather than a causative relationship between stellar activity and measured RVs.

\begin{figure}
	\epsscale{1.2}
	\plotone{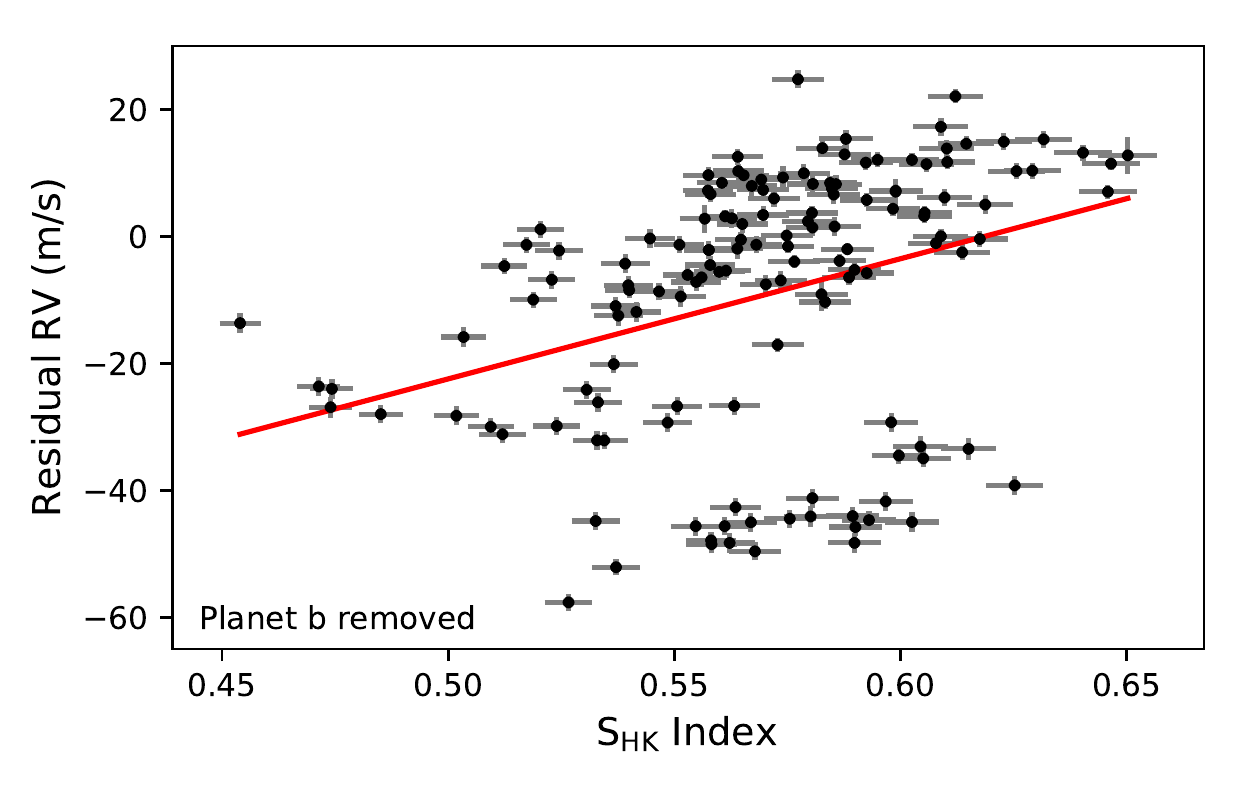}
	\caption{Residual RVs as a function of \Shk index. While there is some correlation between the RVs and stellar activity, the low Pearson's $r$ statistic suggests that only $\sim34\%$ of the total variation could be part of an activity-induced signal.\label{fig:rv_sval_full}}
\end{figure}

\subsection{RV-\Shk Correlation}
\label{ssec:rvsvalcorr}

Finally, if the long-period RV variation was indeed due to stellar activity, they should be correlated across the entire dataset. Figure \ref{fig:rv_sval_full} shows the residual RVs after removing the effect of planet b as a function of \Shk index. While there exists a weak linear correlation, the Pearson's $r$ statistic is only 0.34, indicating that up to a third of the total RMS variation in RVs can be accounted for by stellar activity. The remaining variation, reflected in the large scatter around the fitted line, must be due to another mechanism.

\begin{figure*}
	% \epsscale{1.1}
	\plotone{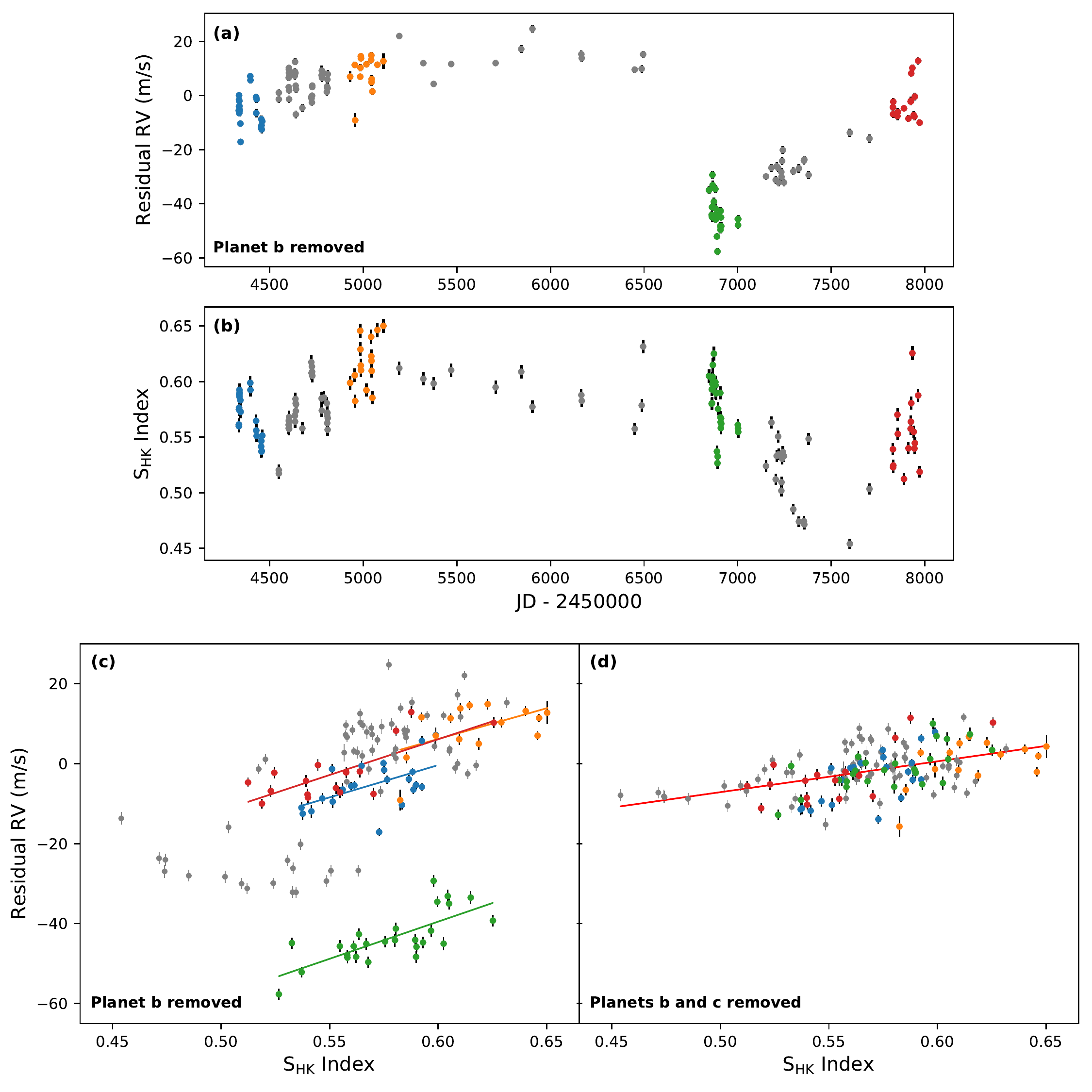}
	\caption{Panel (a): Residual RVs, after removing the effect of the inner planet. We identified four different intervals of up to 180 days each with relatively high observational cadence, over which the RV signal of any putative outer planet can be neglected. These four intervals are marked with colored points.
	Panel (b): \Shk indices, with the same observation periods marked.
	Panel (c): For each identified period, a strong correlation between residual RV and \Shk can be observed, with the slope consistent over all four seasons.
	Panel (d): Same as panel (c) but with the effects of both planets b and c removed. The offsets between the four observing seasons have vanished, such that the correlations from each season are now fully consistent with each other. For clarity, the errors in \Shk index are not shown in panels (c) and (d).
	\label{fig:sval_rv_multi}}
\end{figure*}

\subsection{Summary}
\label{ssec:activity_summary}

Stellar activity alone is insufficient to explain the RV variability of HAT-P-11. The amplitude of the RV variation is too large to be caused solely by stellar activity (Section \ref{ssec:rvmag}), and the activity cycle is offset from the RV signal by more than a year (Section \ref{ssec:phase_mismatch}). The correlation between the residual RVs and \Shk indices also show that most of the RV variation cannot be attributed to stellar activity (Section \ref{ssec:rvsvalcorr}). We therefore subsequently adopt a two-planet interpretation for the data.

\section{RV Modeling}
\label{sec:joint_modeling}

Here, we describe our modeling of the HAT-P-11 RVs that includes contributions from two planets. The weak RV-\Shk correlation, described in Section \ref{ssec:rvsvalcorr}, motivated an analysis that simultaneously includes the effects of stellar activity.

To better understand the connection between the residual RVs and stellar activity, we identify four seasons of $<180$~days with at least 15 observations, over which long-period variations can be neglected (Figure \ref{fig:sval_rv_multi}). For each of these four intervals, we find much stronger linear correlations between the \Shk index and residual RVs than the correlation present in the full set of observations. The Pearson's $r$ statistics were $> 0.5$ for all intervals, and the $p$-values were less than 5\%. We also observe that the high-cadence segments have different mean RVs, but the correlations have consistent slopes as determined by a bootstrap resampling. This suggests that stellar activity has a small but consistent effect on the RV measurements, but it cannot explain the offsets between observing seasons, since they occur at the same \Shk values but have mean RVs that differ by more than 50~\ms.

% \begin{table}
% \centering
% \caption{{\color{red}Correlation between \Shk Index and RVs} \label{tab:period_rv_corr}}
% \begin{tabular}{crrr}
% \hline
% \hline
% \textbf{Interval} & Pearson's $r$	& $p$-value		& Slope \\
% & & & \ms \Shk$^{-1}$ \\
% \hline
% \textbf{1}		& 0.55		& $8.3\times10^{-3}$	& $161\pm30$ \\
% \textbf{2}		& 0.54		& $3.7\times10^{-2}$	& $154\pm47$ \\
% \textbf{3}		& 0.74		& $1.5\times10^{-5}$ 	& $186\pm23$ \\
% \textbf{4}		& 0.75		& $5.1\times10^{-4}$ 	& $179\pm24$ \\
% \hline
% \hline
% \end{tabular}
% \end{table}

Given this short-timescale RV-activity correlation, we modeled the data using a two-planet Keplerian model as well as a linear correlation between the RVs and \Shk. We used the \texttt{RadVel} package \citep{Fulton18} to perform maximum-likelihood fitting and MCMC parameter estimation.

We fixed the period and time of conjunction for HAT-P-11b according to the values derived by \cite{Huber17} from four years of \Kepler data. The remaining orbital parameters for planets b and c, as well as an average RV offset, $\gamma$, were allowed to float. We parameterized $e$ and $\omega$ of each planet as $\sqrt{e}\cos{\omega}$ and $\sqrt{e}\sin{\omega}$ to guard against a bias toward non-zero eccentricities as recommended by \cite{Eastman13}. We also imposed a beta distribution prior for the eccentricities recommended in \cite{Kipping13}. 

The slope of the activity-RV correlation is a new free parameter, $c_S$, such that the induced RV signal is $c_S \Delta\mathrm{S}_{\mathrm{HK,}i}$. Here, $\Delta\mathrm{S}_{\mathrm{HK,}i} \equiv \mathrm{S}_{\mathrm{HK,}i} - \overline{\mathrm{S}_{\mathrm{HK}}}$ is the mean-centered \Shk index at time $t_i$. Any further constant offset is absorbed into $\gamma$.

The likelihood is
$$
\loglike = -\frac{1}{2}\sum_i\left[\frac{(v_i - v_{m,i} - c_S \Delta\mathrm{S}_{\mathrm{HK,}i})^2}{\sigma_i^2+\sigjit{}^2}+\ln 2\pi(\sigma_i^2+\sigjit{}^2)\right]
$$
where $v_i$ and $v_{m,i}$ are the measured and model RVs at time $t_i$, $\sigma_i$ is the corresponding uncertainty on the measured RV, and $\sigma_{\mathrm{jit}}$ is the jitter.

\begin{figure*}
	\epsscale{0.8}
	\plotone{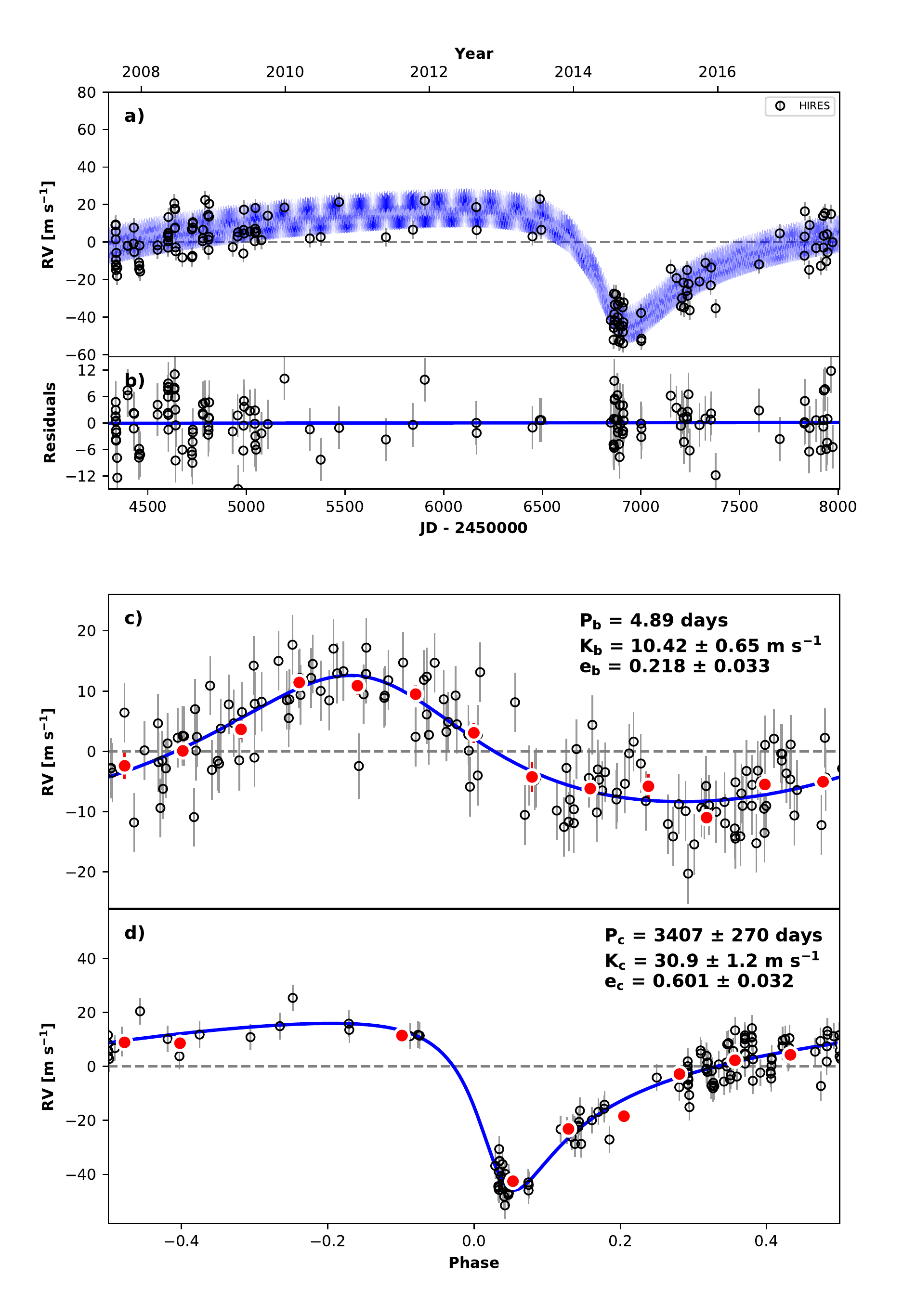}
	\caption{Two-planet Keplerian fit to the RVs, including a linear correlation between \Shk and radial velocities. Panel (a): The most probable model and full radial velocity time series. Panel (b): Residuals from the most probable model, after removing the effect of both planets and the \Shk decorrelation. Panels (c) and (d): Phase-folded RVs and the most probable model for planets b and c respectively, with contributions of the other planet and \Shk decorrelation removed. The large red circles show phase-binned RVs. \label{fig:model_decorr}}
\end{figure*}

The results of our RV fit are shown in Figure \ref{fig:model_decorr} and the derived planetary parameters are given in Table \ref{tab:derived_params}. We also provide the posterior distributions from our MCMC analysis in Appendix \ref{appendix:posteriors}. We find that HAT-P-11c is a $\Mp\sin{i} =$~\val{two-planet-decorr}{M2MJ} $M_J$ giant planet with semimajor axis of $a=$~\val{two-planet-decorr}{a2}~AU. Its high-eccentricity orbit ($e = $~\val{two-planet-decorr}{e2}) gives it a periastron distance of \val{two-planet-decorr}{rp2}~AU and an apoastron distance of \val{two-planet-decorr}{ra2}~AU. This large separation reached at apoastron will have a positive effect on any future attempts to detect the planet via direct imaging, as we discuss in Section \ref{sec:future_obs}.

\begin{table}
\centering
\caption{System Parameters \label{tab:derived_params}}
\begin{tabular}{lrr}
\hline
\hline
\multicolumn{2}{l}{\textbf{Stellar Parameters}}	&  \\
\hline
\Rstar (\Rsun) & \val{stellar}{radius}		& A 	\\
\Mstar (\Msun) & \val{stellar}{mass}		& B 	\\
\teff (K) 	   & \val{stellar}{teff}		& B		\\
\fe 		   & \val{stellar}{fe}			& B		\\
$V$ (mag)	   & \val{stellar}{Vmag}		& B		\\
\vsini  (\kms) & \val{stellar}{vsini}	& B	\\
$P_\mathrm{rot}$ (d) & \val{stellar}{prot}	& B		\\
Age (Gyr) 	   & \val{stellar}{age}			& B		\\
Distance (pc)  & \val{stellar}{distance}	& C  	\\
\hline
\multicolumn{3}{l}{\textbf{Planetary Parameters}}	\\
\textbf{Planet b} & & \\
\hline
$P$ (days)		& \val{two-planet-decorr}{P1}	 		& D \\
$T_{\rm{conj}}$ (JD)  & \val{two-planet-decorr}{T01}	 	& D \\
$e$ 				& \val{two-planet-decorr}{e1} 			& E \\
$\omega$ (\deg)		& \val{two-planet-decorr}{w1} & E \\
$M_P \sin{i}$ (\Me) & \val{two-planet-decorr}{M1} 			& E \\
$a$ (\AU) 			& \val{two-planet-decorr}{a1} 			& E \\
$R_{\mathrm{P}}$ (\Re)	& $4.36 \pm 0.06$ 	  				& D \\
$r_{\rm{peri}}$ (AU) & \val{two-planet-decorr}{rp1} 		& E \\
$T_{\rm{peri}}$ (JD) & \val{two-planet-decorr}{tp1} 		& E \\
$r_{\rm{apo}}$ (AU) & \val{two-planet-decorr}{ra1} 			& E \\
$T_{\rm{apo}}$ (JD) & \val{two-planet-decorr}{ta1} 			& E \\
\hline
\textbf{Planet c} & & \\
\hline
$P$ (days)			& \val{two-planet-decorr}{P2} 	& E \\
$T_{\rm{conj}}$ (JD)  	& \val{two-planet-decorr}{T02} 	& E \\
$e$ 					& \val{two-planet-decorr}{e2} 	& E \\
$\omega$ (\deg)			& \val{two-planet-decorr}{w2} 	& E \\
$M_P \sin{i}$ (\Me) 	& \val{two-planet-decorr}{M2} 	& E \\
$a$ (\AU) 				& \val{two-planet-decorr}{a2} 	& E \\
$r_{\rm{peri}}$ (AU) 	& \val{two-planet-decorr}{rp2} 	& E \\
$T_{\rm{peri}}$ (JD) 	& \val{two-planet-decorr}{tp2} 	& E \\
$r_{\rm{apo}}$ (AU) 	& \val{two-planet-decorr}{ra2} 	& E \\
$T_{\rm{apo}}$ (JD) 	& \val{two-planet-decorr}{ta2} 	& E \\
\hline
\multicolumn{3}{l}{
\textbf{A:} \cite{Deming11}
\textbf{B:} \cite{Bakos10}
} \\
\multicolumn{3}{l}{
\textbf{C:} \cite{Gaia16a}
} \\
\multicolumn{3}{l}{
\textbf{D:} \cite{Huber17}
\textbf{E:} This work
} \\
\end{tabular}
\end{table}

Once the effect of both planets is removed (Figure \ref{fig:sval_rv_multi}d), residual RVs show a strong linear correlation with the \Shk values (Pearson's $r=0.479$, $p$-value~$=8\times10^{-9}$), where the offsets between the four observing seasons are eliminated. The total semi-amplitude of the activity-induced RV is $\sim7$~\ms, consistent with that observed in stars of similar spectral types (see Section \ref{ssec:rvmag}).

We also investigated models with higher and lower complexity. We first examined a single-planet model with activity as well as a two-planet model without activity correction. These models were not favored when compared using the Bayesian Information Criterion (BIC; \citealt{Schwartz78}). We also considered the possibility of additional planets in the system, but these were not found by a two-dimensional Keplerian Lomb-Scargle (2DKLS; \citealt{OToole09}) periodogram search. We describe these model comparisons in detail in Appendix \ref{appendix:model_comparison}.

Finally, to ensure our methodology does not always favor planets over activity, we applied an identical analysis to the HD99492 system, another active mid-K dwarf with long-period activity and RV signals (Appendix \ref{appendix:hd99492}). In this case, the BIC rejects a planetary explanation for the RVs and prefers a pure stellar activity model, in agreement with the findings of \cite{Kane16}.

\section{System Dynamics and Spin-Orbit Misalignment}
\label{sec:dynamics}

The orbit of HAT-P-11b is known to be misaligned with its host star's spin axis, with an obliquity of $\lambda \approx 100\deg$, corresponding to a nearly polar orbit (see Section \ref{sec:intro}). There are a number of other planets with misaligned orbits (see \citealt{Albrecht12b,Dai17}). Many explanations have been proposed for such misalignments, including Kozai-Lidov cycles (e.g. \citealt{Fabrycky07}), planet-planet scattering (e.g. \citealt{Nagasawa08}), primordial tilting of the protoplanetary disk (e.g. \citealt{Batygin12}), or angular momentum transport by internal gravity waves (e.g. \citealt{Rogers12}). Here, we examine the dynamical coupling between HAT-P-11b and c and assess if it can explain the observed misalignment.

The orbital angular momentum of HAT-P-11c is much greater than that of HAT-P-11b and the star's spin angular momentum, allowing us to make the approximation that the orbital plane of planet c is invariant. As a matter of convenience, we define angles that describe the orientation of planet b's orbit, inclination $i_b$ and argument of periastron $\omega_b$, with respect to the orbital plane of planet c. Note that this reference plane is not the sky plane, which is often used to describe the orbits of transiting planets. In this coordinate system, $i_b$ is therefore the relative inclination between the two planets. Following \cite{Mardling10}, we write down the orbit-averaged Hamiltonian for the interaction of the two planets, expanded to quadrupole order in semimajor axis ratio \citep{Kaula64}:\footnote{The semimajor axis ratio in this sytem is $\frac{a_b}{a_c} \approx 0.01$, warranting a leading-order trunctation of the Hamiltonian.}
\begin{equation} \label{eq:or_hamiltonian}
\begin{aligned}
\mathcal{H} = & \frac{1}{4}\frac{\mathcal{G} m_b m_c}{a_c} \left(\frac{a_b}{a_c}\right)^2 \left(\frac{1}{\sqrt{1-e_c^2}}\right)^3 \\
& \left[\left(1 + \frac{3}{2}e_b^2 \right) \left(\frac{3}{2} \cos^2{i_b} - \frac{1}{2}\right) + \frac{15}{4} e_b^2 \sin^2{i_b} \cos{2\omega_b}\right].
\end{aligned}
\end{equation}
The second term, containing $e_b^2 \sin^2{i_b}$, gives rise to the Kozai-Lidov mechanism, in which the inner planet undergoes cycles trading large inclinations for large eccentricities. However, because HAT-P-11b is very close in to its star, general relativistic (GR) effects cause apsidal precession, which may suppress Kozai-Lidov oscillations under suitable conditions (e.g. \citealt{Ford00,Fabrycky07}). Hence, we must include an additional GR term in the Hamiltonian.

We can write the Hamiltonian including the GR term using scaled canonical Delaunay variables, where $\Omega_b$ is the longitude of ascending node of planet b:
\begin{equation*} 
\begin{aligned}
G = \sqrt{1-e_b^2} 			 \qquad &g = \omega_b \\
H = \sqrt{1-e_b^2} \cos{i_b} \qquad &h = \Omega_b
\end{aligned}
\end{equation*}
and correspondingly scaling the Hamiltonian by $m_b \sqrt{\mathcal{G}M_{\star}a_b}$, giving
\begin{equation} \label{eq:hamiltonian}
\begin{aligned}
\mathcal{H}^{\prime} & = \frac{1}{16}n_b \frac{m_c}{M_{\star}}\left(\frac{a_b}{a_c\sqrt{1-e_c^2}}\right)^3 
\left[\frac{\left(5 - 3G^2\right)\left(3H^2 - G^2\right)}{G^2} \right. & \\
& \left. + \frac{15 \left(1 - G^2\right) \left(G^2 - H^2 \right) \cos{2g}}{G^2} \right] + 3 n_b \frac{\mathcal{G} M_{\star}}{a_b c^2}\frac{1}{G}.
\end{aligned}
\end{equation}
Here, we have written the expression in terms of the mean motion $n_b = \sqrt{\mathcal{G}M_{\star}/a^3}$.

The rapid apsidal precession due to GR may suppress the Kozai resonance, which requires a slowly varying $\omega_b$. We calculate the GR precession rate,
\begin{equation*}
\begin{aligned}
\dot{\omega}_{GR} & = 3n_b \frac{\mathcal{G}M_{\star}}{a_b c^2}\frac{1}{G^2}\\
& \approx 2.2\times10^{-4} \,\rm{yr}^{-1}
\end{aligned}
\end{equation*}
which gives a precession period of approximately 30,000 years.

In comparison, the Kozai timescale is given by \citep{Kiseleva98}:
\begin{equation*}
\begin{aligned}
\tau & = \frac{2P_c^2}{3\pi P_b^2} \frac{M_\star}{m_c}\left(1 - e_c^2\right)^{3/2}\\
& \approx 4\times10^5 \,\rm{years},
\end{aligned}
\end{equation*}
an order of magnitude longer. Thus, we expect that the Kozai mechanism is suppressed in this system.

To confirm this, we examine the phase space of the Hamiltonian (\ref{eq:hamiltonian}). The Hamiltonian admits two integrals of motion: $H$ as well as $\mathcal{H}^{\prime}$ itself. Thus, any given phase-space portrait is parameterized by $H$, which translates to a particular $i_{\rm{max}}$, the inclination of the inner planet attained when its orbit is circular. Along level curves of $\mathcal{H}^{\prime}$, the variables $G, g$ trace out trajectories in a two-dimensional phase space, where the eccentricity is given by $e_b = \sqrt{1 - G^2}$, which specifies the instantaneous inclination via the conservation of $H$.

\begin{figure*}
	\epsscale{0.9}
	\plotone{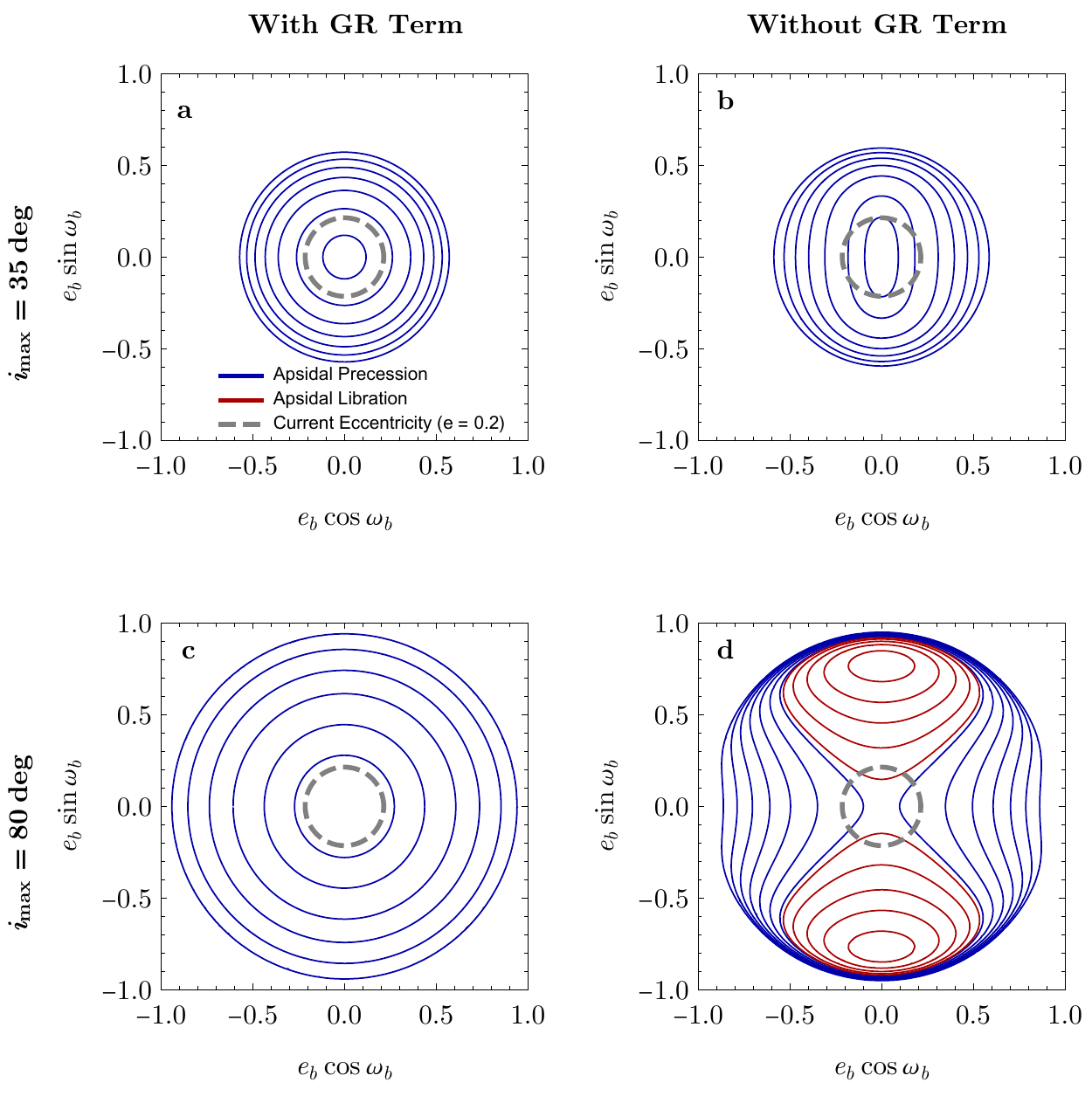}
	\caption{Phase-space portraits for the two-planet Hamilton in $e_b$, $\omega_b$ space for different values of $i_{\rm{max}}$. Blue trajectories indicate circulation, where $e$ remains roughly constant while $\omega$ precesses; red trajectories indicate libration of $e$. The dashed gray circle shows the current observed eccentricity of HAT-P-11b. 
	Panels (a), (b): At low maximal inclinations $i_{\rm{max}}$, only circulatory trajectories exist, where $e_b$ remains roughly constant while $\omega_b$ precesses. 
	Panel (c): GR precession suppresses libratory trajectories even at high values of $i_{\rm{max}}$.
	Panel (d): If GR precession is neglected, the Kozai mechanism can occur, with libratory trajectories taking the planet to high eccentricity and inclination. \label{fig:phase_portraits}}
\end{figure*}

We plot the phase-space portraits for two different values of $i_{\rm{max}}$ with and without GR, projected into non-canonical coordinates $e_b\cos{\omega_b}$ and $e_b\sin{\omega_b}$ in Figure \ref{fig:phase_portraits}. In agreement with the simple timescale argument presented above, we find that the fast precession of $\omega_b$ in the HAT-P-11 system is sufficient to suppress Kozai oscillations, such that there is no libration of $e_b$ for any value of $i_{\rm{max}}$.

We note that simply because the Kozai effect does not operate within the present-day architecture of the HAT-P-11 system does not rule out the possibility that it could have operated previously. This requires the semimajor axis of planet b to have been larger in the past and to have shrunk to its current configuration due to tidal friction (e.g. \citealt{Fabrycky07}). However, the tidal migration timescale is longer than the tidal circularization timescale by a factor of $1/(1-e^2)$ \citep{Hut81}. Thus, tides would tend to circularize the orbit faster than they shrink the orbit. Given that the orbit is still eccentric, we consider it unlikely that tidal damping has shrunk the orbit of HAT-P-11b.

\begin{figure}
	% \epsscale{1.2}
	\plotone{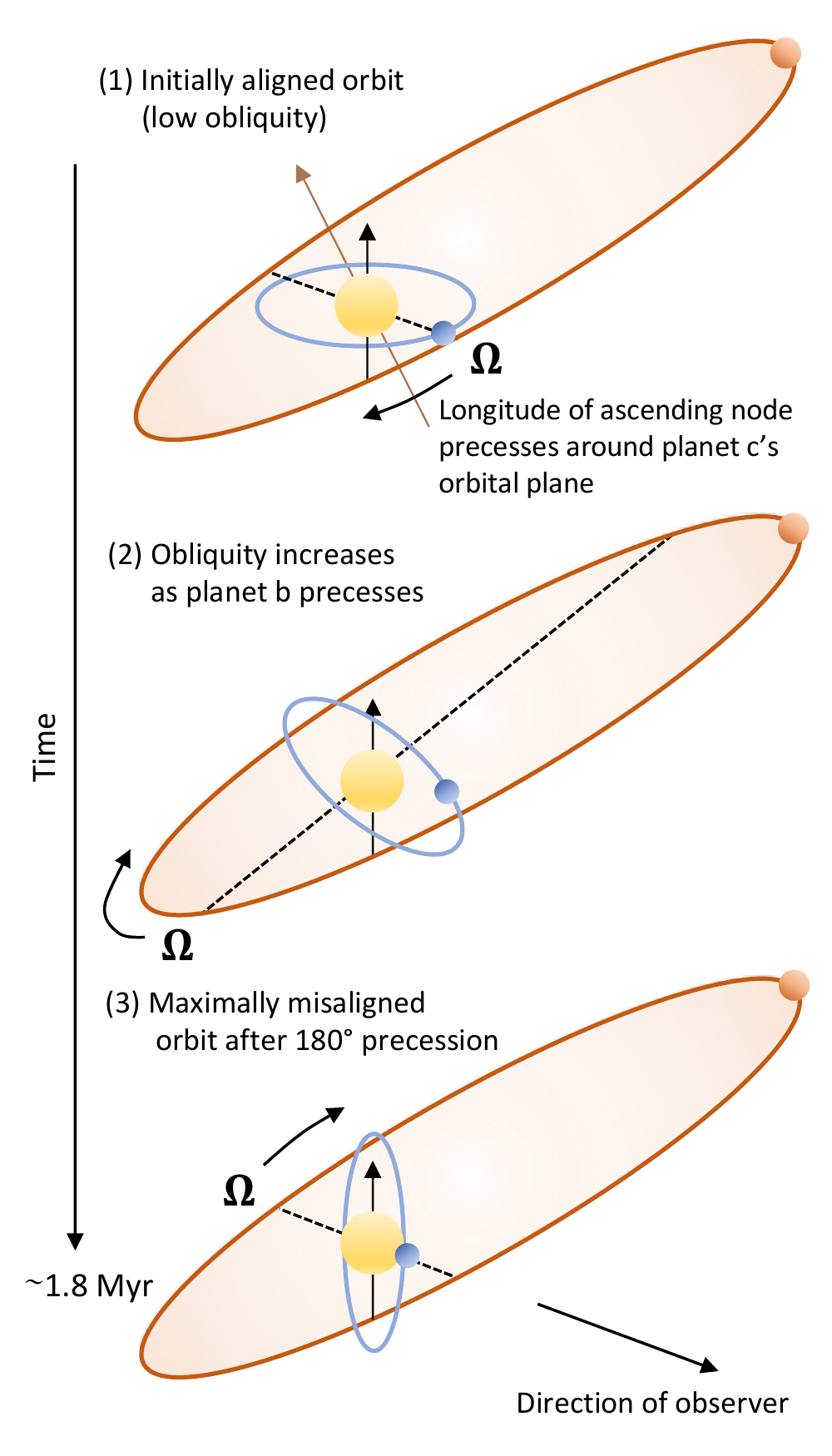}
	\caption{Precession of the longitude of ascending node $\Omega_b$ of HAT-P-11b's orbit (blue) around the orbital plane of HAT-P-11c (brown) can result in a polar orbit for planet b. (1) Initially, the inner planet's orbit is aligned with the star's rotation axis (black arrow), but the outer planet has an inclination $\sim 50\deg$ (normal to orbital plane shown as brown arrow). (2) As time progresses, the longitude of ascending node precesses around the plane of the outer planet's orbit, increasing the inclination of the inner planet's orbit relative to the stellar rotation axis. (3) The inner planet reaches a maximum inclination twice that of the outer planet relative to the stellar rotation axis.  \label{fig:orbit_precession}}
\end{figure}

Given the lack of Kozai cycles, we can average out the Kozai term, which leaves us with a trivial dynamical system, governed by a Hamiltonian that only depends on the actions, and thus yields only precession as its consequence. The longitude of ascending node then evolves according to
\begin{equation} \label{eq:prec}
\begin{aligned}
\frac{d\Omega_b}{dt} &= \frac{\partial \mathcal{H}'}{\partial H} \\
&= \frac{1}{2}n_b\frac{m_c}{M_{\star}}\left(\frac{a_b}{a_c\sqrt{1-e_c^2}}\right)^3\left(\frac{15 - 9G^2}{G^2}\right)H.
\end{aligned}
\end{equation}
Thus, $\Omega_b$ precesses around the invariant plane defined by the outer planet's orbit, with a period of approximately 3.5 Myr, significantly shorter than the age of the system. If the orbit normal of planet c is misaligned with the spin axis of the star by more than half the current observed obliquity of planet b, $i_c \gtrsim 50\deg$,  this would be sufficient to explain HAT-P-11b's approximately polar orbit (Figure \ref{fig:orbit_precession}). However, this does not explain the initial misalignment of HAT-P-11c, which may be the result of planet-planet scattering in the outer system.

We also check that the stellar spin axis does not itself precess more quickly than the inner planet, as such an arrangement would result in a coupling between the star and the inner planet, allowing both to precess together and remain aligned. We follow \cite{Spalding15} and model the oblateness of the star as a point mass surrounded by an orbiting ring with effective semimajor axis
\begin{equation}
\begin{aligned}
\tilde{a} &= \left[\frac{16 \nu^2 k_2^2 R_{\star}^6}{9 I^2 \mathcal{G} M_{\star}}\right]^{1/3}.
\end{aligned}
\end{equation}
Here, $\nu$ is the stellar rotation frequency, and we have used the dimensionless moment of inertia $I = 0.08$ and Love number $k_2 = 0.01$. Using Equation \ref{eq:prec} for the precession rate due to the torque from the inner planet on the star, we confirm that the stellar precession timescale is on the order of 100 Myr, much slower than the precession of the inner planet.

This application of secular theory to the HAT-P-11 system presents a plausible dynamical history that explains the unusual polar orbit of HAT-P-11b. Through precession around the outer planet's orbital plane, HAT-P-11b can attain very high obliquities with respect to the stellar rotation axis, although the angle with respect to the invariant plane remains fixed. A measurement of the mutual inclination between the planetary orbits, for example, through astrometry (Section \ref{sec:future_obs}), could help shed more light on this explanation. Nonetheless, irrespective of the exact scenario, this system would have required a large degree of primordial misalignment, either between the orbits of the two planets as described here, or between the stellar spin axis and HAT-P-11b.

\section{The HAT-P-11 System in Context}
\label{sec:context}

\begin{figure*}
	\epsscale{1.2}
	\plotone{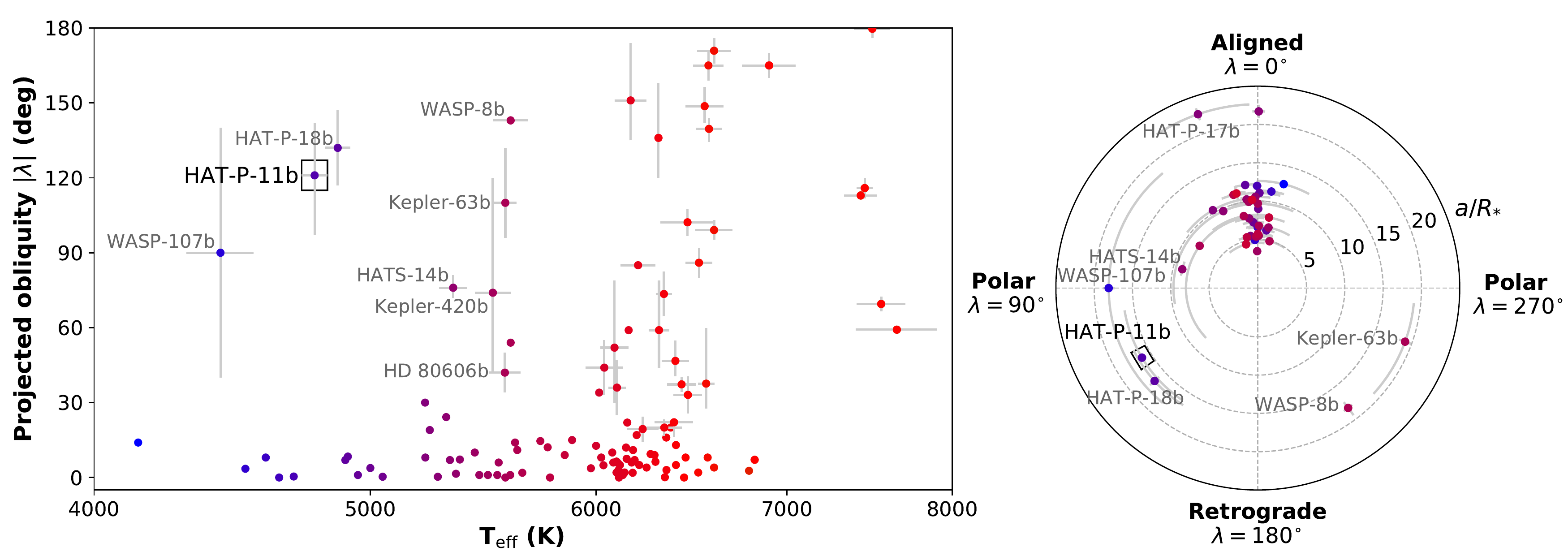}
	\caption{HAT-P-11b in the context of other planets with measured obliquities.\footnotemark[3] Left: Effective temperature and projected stellar obliquities. For clarity, error bars are only shown for planets misaligned by $10\deg$ or more. A large fraction of planets orbiting stars hotter than $\sim6000$~K are misaligned, perhaps due to the absence of convective envelopes around these stars (see Section \ref{sec:context}). HAT-P-11b is one of only a handful of misaligned planets orbiting a star cooler than 6000~K. Right: Obliquities for stars with \teff~$< 6000$~K as a function of $a / \Rstar$. Close-in planets tend to be aligned, but this preference vanishes for $a / \Rstar \gtrsim 15$. The mapping between point color and \teff is the same as in the left panel. \label{fig:obliquity_corr}}
\end{figure*}

Among the planets that have measured obliquities, HAT-P-11b is an outlier. It has the smallest planetary to stellar mass ratio and one of the lowest host star effective temperatures for a misaligned planet (Figure \ref{fig:obliquity_corr}).

\cite{Winn10b} first noted a connection between high stellar obliquities and effective temperature, with a higher proportion of planets around hot stars (\teff~$\gtrsim 6000$~K) with misaligned orbits. They suggested that this could be due to the fact that cool stars have larger convective zones, creating strong tidal coupling with their close-in planets that realigns the star to the planet's orbit normal. In contrast, hot stars without these large convective zones have weaker tidal coupling, leading to a longer tidal realignment timescales.

Indeed, among the cool stars (\teff~$< 6000$~K) with obliquity measurements, most of the systems exhibiting significant misalignment also have large $a / \Rstar$ (Figure \ref{fig:obliquity_corr}). This corresponds to a longer realignment timescale, allowing systems to retain any primordial inclination. Seen in this light, HAT-P-11b is no longer an outlier, with $a / \Rstar = 16.3 \pm 0.4$. \cite{Albrecht12b} calculated the characteristic realignment timescale for the system to be $\sim 10^{15}$ years, vastly longer than the age of the system. Hence, while the secular precession of HAT-P-11b's orbit normal may be a plausible reason for the observed misalignment, it is not a necessary condition since any primordial misalignment of HAT-P-11b would have also been retained.

While nodal precession is not necessary to maintain HAT-P-11b's misalignment for Gyr timescales, it may be for shorter-period planets like HATS-14b \citep{Zhou15}. Another system for which this mechanism may be at work is WASP-8, which is reminiscent of HAT-P-11 in the following respects: WASP-8 is a cool star with a close-in misaligned planet and a distant giant planet on an eccentric orbit \citep{Knutson14}. 

Nodal precession is likely one of several ways to produce misaligned planets. For example, Kepler-420b \citep{Santerne14} and HD 80606b \citep{Hebrard10} have stellar companions, suggesting star-planet rather than planet-planet mechanisms. However, if planet-planet nodal precession is a common mechanism to produce misaligned orbits, we predict a correlation between misaligned planets around cool stars and distant, eccentric giants. Such a prediction is testable with future RV, astrometric, or imaging follow-up of known misaligned systems as well as the many more that will soon be discovered by the {\em Transiting Exoplanet Survey Satellite} (TESS; \citealt{Ricker14}).

\footnotetext[3]{Data compiled from TEPCat as of October 2017 (\citealt{Southworth11}, \url{http://www.astro.keele.ac.uk/jkt/tepcat/rossiter.html}).}

\section{Prospects for Future Observations}
\label{sec:future_obs}

\subsection{Secondary Eclipse of HAT-P-11b}
\label{ssec:secondary_eclipse}

Secondary eclipse observations can provide insight into the albedo and thermal structure of a planet's atmosphere. \cite{Huber17} reported the detection of the secondary eclipse of HAT-P-11b based on \Kepler photometry, at an orbital phase of $\phi = 0.659$ and depth of 5~ppm. Our adopted RV model found an expected secondary eclipse phase of $\phi=0.623^{+0.018}_{-0.019}$, consistent within 2$\sigma$ of the \cite{Huber17} detection. Future observations with the {\em James Webb Space Telescope} \citep{Gardner06} should be able to detect and characterize this secondary eclipse.

\subsection{Astrometric Characterization of HAT-P-11c}
\label{ssec:astrometry}

The \Gaia mission is currently making extremely high-precision astrometric observations of one billion astronomical objects. For HAT-P-11, \Gaia is expected to reach an astrometric precision of $\sim 7 \uas$ \citep{Gaia16b} by the end of its five-year nominal mission.

The wide orbit ($a=$~\val{two-planet-decorr}{a2}~AU) of HAT-P-11c and the system's relative proximity (\val{stellar}{distance}~pc) to Earth means that HAT-P-11c will yield a maximum astrometric signal of \val{two-planet-decorr}{astrom}~\uas. The presence of the planet should in principle be detectable by \Gaia during its nominal five-year mission, while a full determination of the orbit's three-dimensional orientation will be possible if the mission duration is extended to 10~years. Although such a measurement will not uniquely determine the mutual inclination of the two planets, it will constrain the difference in angles in a single plane and could help verify the dynamical picture described in Section \ref{sec:dynamics}. 

\subsection{Direct Imaging of HAT-P-11c}
\label{ssec:direct_imaging}

HAT-P-11c's eccentric orbit takes it up to \val{two-planet-decorr}{ra2}~AU from its host star. Taking into account the argument of periastron of the orbit, the maximum sky-projected planet-star separation is \val{two-planet-decorr}{angsep2}~mas. 

Assuming a radius and albedo similar to that of Jupiter, the reflected-light contrast ratio of the planet will be $\sim6\times10^{-9}$. This makes HAT-P-11c a potential although challenging target for high-contrast direct imaging studies. For example, the {\em Wide Field Infrared Survey Telescope} (WFIRST, \citealt{Wfirst15}) is expected to have a $10^{-9}$ effective contrast and 100 mas inner working angle, and may be able to characterize HAT-P-11c.

\section{Conclusion}
\label{sec:conclusions}

The HAT-P-11 system is one of the best-studied exoplanet systems, with long baseline RV and photometric datasets and RM measurements. Here, we extend the RV baseline to ten years and discover a new $1.5\,M_J$ giant plant on a distant, eccentric orbit. We found that the presence of HAT-P-11c may explain the previously known misalignment of HAT-P-11b through nodal precession. This mechanism may help to explain the diversity of exoplanet obliquities as a function of orbital distance and host star type. Further characterization of the HAT-P-11 system will soon be possible thanks to upcoming spectroscopic, astrometric, and imaging facilities.

\acknowledgments 
We thank Joshua Winn for helpful discussions that improved the final manuscript. 
We thank the many observers who contributed to the measurements reported here including M.~Bottom, B.~Bowler, P.~Butler, J.~Brewer, J.~Crepp, C.~Chubak, K.~Clubb, I.~Crossfield, D.~Fischer, E.~Ford,  E.~Gaidos, M.~Giguere, J.~Johnson, M.~Kao, G.~Marcy, T.~Morton, G.~Mandushev, K.~Peek, S.~Pineda, G.~Torres, S.~Vogt, P.~Worden, M.~Zhao. 

S.~W.~Y.\ acknowledges support from the Caltech Marcella Bonsall Summer Undergraduate Research Fellowship.
E.~A.~P.\ acknowledges support from a Hubble Fellowship grant HST-HF2-51365.001-A awarded by the Space Telescope Science Institute, which is operated by the Association of Universities for Research in Astronomy, Inc. for NASA under contract NAS 5-26555. 
The data presented herein were obtained at the W. M. Keck Observatory, which is operated as a scientific partnership among the California Institute of Technology, the University of California and the National Aeronautics and Space Administration. The Observatory was made possible by the generous financial support of the W. M. Keck Foundation.
The authors wish to recognize and acknowledge the very significant cultural role and reverence that the summit of Maunakea has long had within the indigenous Hawaiian community.  We are most fortunate to have the opportunity to conduct observations from this mountain.
\software{\texttt{Numpy/Scipy} \citep{VanDerWalt11}, \texttt{Matplotlib} \citep{Hunter07}, \texttt{Pandas} \citep{McKinney10}, \texttt{Astropy} \citep{Astropy13}, \texttt{emcee} \citep{Goodman10,Foreman-Mackey13}, \texttt{RadVel} \citep{Radvel18}}

\facility{Keck:I (HIRES)}

%\vspace{5mm}
% \nocite{apsrev41Control}
\bibliography{revtex-custom,manuscript,software}

\appendix

\section{Model Comparison and Selection}
\label{appendix:model_comparison}

In Section \ref{sec:joint_modeling}, we adopted an RV model incorporating the effect of two planets and an activity-induced signal. This model was chosen after exploring two other possibilities: (1) A single-planet model where the long-term RV trend must be completely accounted for by the activity-RV correlation, and (2) A two-planet model without any activity-RV correlation. Model (3) is the two-planet plus activity model. To compare the models, we compute the BIC, which incorporates the log-likelihood of the model and a penalty for the number of free parameters. A model with lower BIC is preferred, with $|\dbic| \gtrsim 10$ being strongly favored. We present the results of the RV fits in Table \ref{tab:rv_fit_results}.

We found that the single-planet model (1) gave the poorest fit, with RMS residuals of almost 17 \ms and a high BIC. Models (2) and (3) provided a significant improvement in the RMS residuals, indicating that the long-period signal can be best explained by the presence of an outer planet. The model parameters found by both models for the two planets are similar, typically differing by less than 1-$\sigma$. However, model (3) had the lowest BIC ($\dbic = -16$) despite the additional model complexity. This is in accordance with our conclusions from Section \ref{sec:planet_evidence}, where we saw that stellar activity does account for some, but not all, of the RV variation. The BIC analysis thus supports our choice of a two-planet RV model with a linear activity-RV correction.

\begin{table*}[ht]
\fontsize{6}{7.2}
\centering
\caption{Comparison of RV-Activity Models \label{tab:rv_fit_results}}
\begin{tabular}{lrrr}
\hline
\hline
& \textbf{Model 1} 	& \textbf{Model 2} & \textbf{Model 3} \\
& & & \textbf{(Adopted)} \\
\hline
Number of planets 	 	& 1		& 2		& 2		\\
\Shk correction			& Yes 	& No 	& Yes 	\\
Number of free parameters 		& 6		& 10 	& 11	\\
\hline
\multicolumn{4}{l}{\textbf{Planet b parameters}} \\
$P_{b}$ (days)			& \multicolumn{3}{c}{\val{single-planet-model}{P1}} 	\\
$T\rm{conj}_b$ (JD)		& \multicolumn{3}{c}{\val{single-planet-model}{T01}}	\\
$e_{b}$ 				& \val{single-planet-model}{e1} & \val{two-planet-model}{e1} & \val{two-planet-decorr}{e1} \\
$\omega_{b}$ (\deg)		& \val{single-planet-model}{w1} & \val{two-planet-model}{w1} & \val{two-planet-decorr}{w1} \\
$K_{b}$ (\ms) 			& \val{single-planet-model}{K1} & \val{two-planet-model}{K1} & \val{two-planet-decorr}{K1} \\
\hline
\multicolumn{4}{l}{\textbf{Planet c parameters}} \\
$P_{c}$ (days)			& - 							& \val{two-planet-model}{P2} & \val{two-planet-decorr}{P2} \\
$T\rm{conj}_c$ (JD)		& - 							& \val{two-planet-model}{T02} & \val{two-planet-decorr}{T02} \\
$e_{c}$ 				& - 							& \val{two-planet-model}{e2} & \val{two-planet-decorr}{e2} \\
$\omega_{c}$ (\deg)		& - 							& \val{two-planet-model}{w2} & \val{two-planet-decorr}{w2} \\
$K_{c}$ (\ms) 			& - 							& \val{two-planet-model}{K2} & \val{two-planet-decorr}{K2} \\
\hline
\multicolumn{4}{l}{\textbf{Global parameters}} \\
$c_S$ (\ms \Shk$^{-1}$) & \val{single-planet-model}{c1} & - 						 & \val{two-planet-decorr}{c1} \\
$\gamma$ (\ms)			& \val{single-planet-model}{gamma} & \val{two-planet-model}{e1} & \val{two-planet-decorr}{gamma} \\
\sigjit{} (\ms)			& \val{single-planet-model}{jit} & \val{two-planet-model}{jit} & \val{two-planet-decorr}{jit} \\
\hline
\multicolumn{4}{l}{\textbf{Model comparison}} \\
RMS residuals (\ms)		& \val{single-planet-model}{rms} & \val{two-planet-model}{rms} & \val{two-planet-decorr}{rms} \\
BIC						& \val{single-planet-model}{bic} & \val{two-planet-model}{bic} & \val{two-planet-decorr}{bic} \\
\hline
\hline
\end{tabular}
\tablecomments{$c_S$ is the slope of the linear correlation between RV and \Shk}
\end{table*}

We also investigated more complex models to determine if there were additional planets in the system. We performed an iterative search using the two-dimensional Keplerian Lomb-Scargle (2DKLS) periodogram \citep{OToole09} following the technique of \cite{Fulton15} and \cite{Howard16}. An empirical False Alarm Probability (eFAP) is calculated from a histogram of periodogram amplitudes.

\begin{figure}[h]
	\plotone{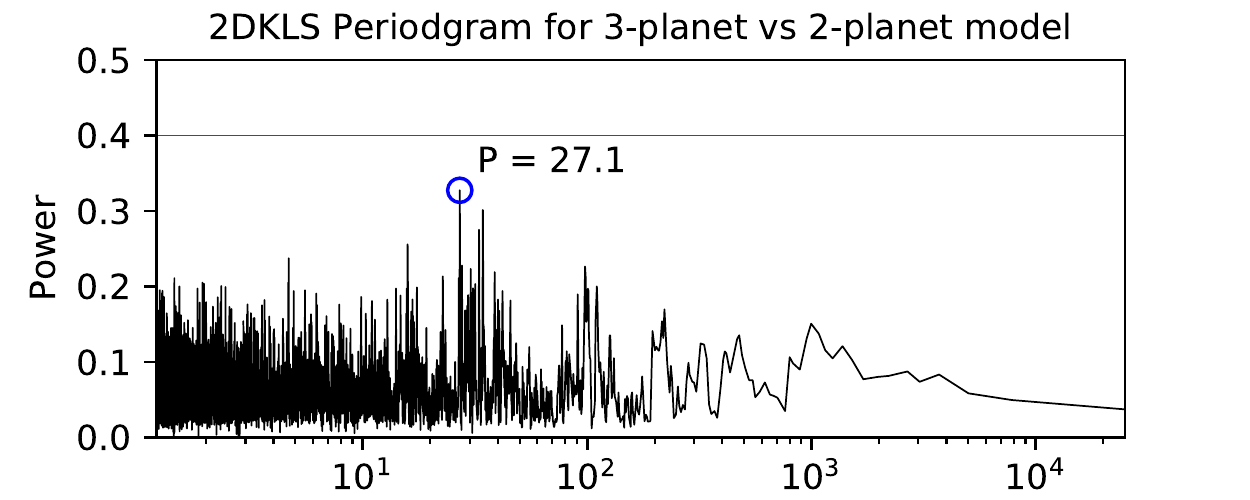}
	\epsscale{0.7}
	\caption{Two-Dimensional Keplerian Lomb-Scargle (2DKLS) periogram where a third Keplerian is fit to the residuals after removing HAT-P-11b and c. No additional significant periodic signals could be detected. \label{fig:2dkls}}
\end{figure}

Figure \ref{fig:2dkls} shows the periodogram for a three- versus two-planet model. We find no significant peaks above the eFAP threshold of 1\%, with only minor peaks close to the stellar rotation period of 29 days and its aliases. Thus, inclusion of a third planet is not justified given the current dataset.

\section{Comparison with HD 99492}
\label{appendix:hd99492}

To validate the joint RV-activity methodology described in section \ref{sec:joint_modeling}, we applied the same analysis to the HD~99492 system, another moderately active mid-K dwarf with a decade of RV measurements. Similar to HAT-P-11, HD~99492 has a relatively short-period ($P=17.1$ days) inner planet first discovered by \cite{Marcy05}. \cite{Meschiari11} later attributed a 5000~day RV signal to a distant giant planet. However, the star's activity cycle, as traced by the \Shk index, was found to have similar periodicity by \cite{Kane16}, who subtracted the effect of the inner planet from the RVs and found a strong correlation between these residual RVs and the \Shk values. They used this to argue that the outer planet reported by \cite{Meschiari11} should therefore be attributed to stellar activity. 

\begin{figure}[h]
	\epsscale{0.5}
	\plotone{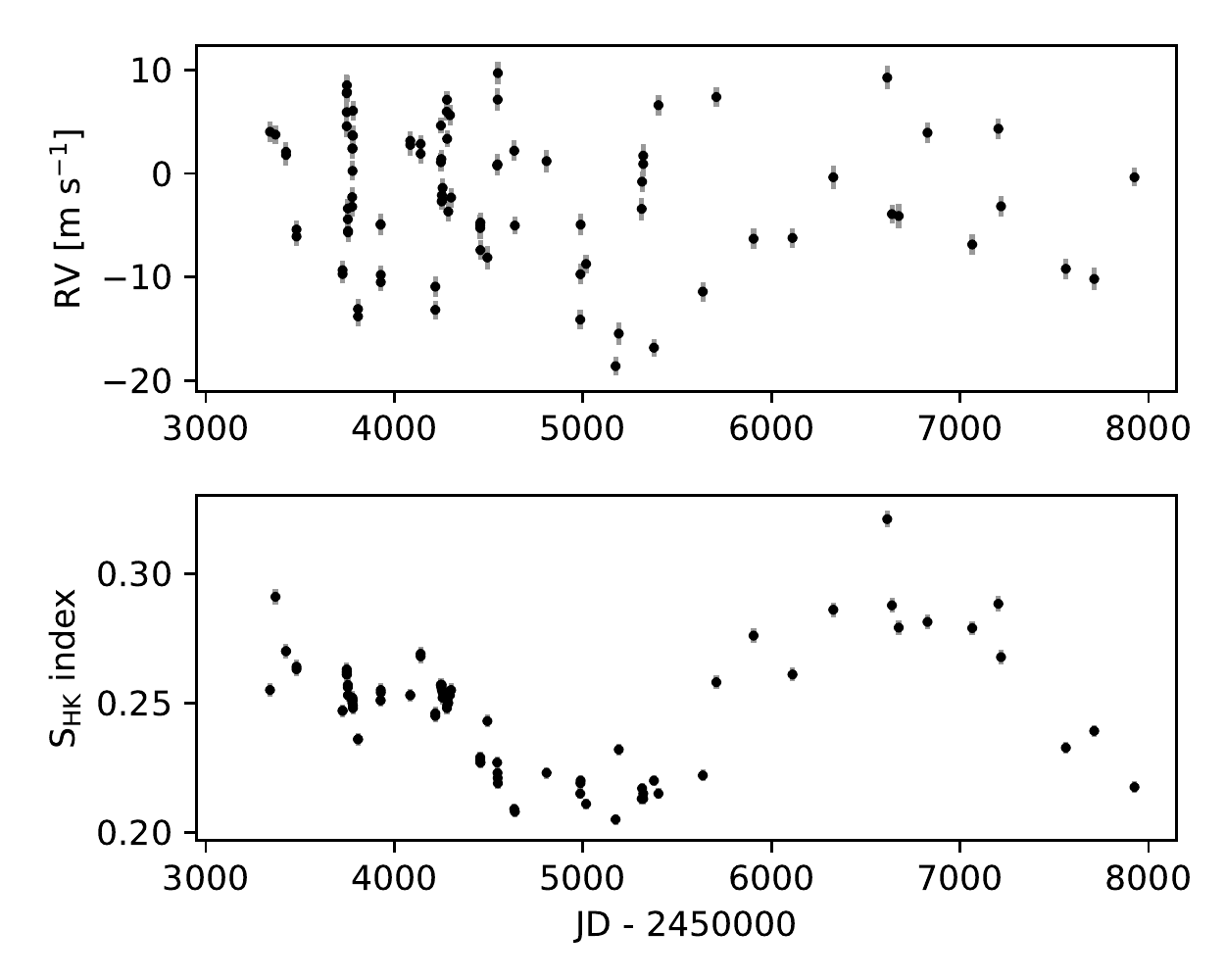}
	\caption{RV measurements (top) and \Shk indices (bottom) for HD 99492. Similar to HAT-P-11, the \Shk index exhibits a long-term variation indicative of an activity cycle. \label{fig:hd99492_timeseries}}
\end{figure}

We investigated the HD~99492 system using the same joint activity-RV analysis described previously. We used 89 HIRES spectra taken by the CPS program over 13 years, from 2004 to 2017 (Figure \ref{fig:hd99492_timeseries}). A long-term periodic signal can clearly be seen in the \Shk measurements, of almost identical period and phase to the RV variability. 

We then fit these RV measurements using three different models, similar to those used for fitting HAT-P-11: (1) A single-planet model with activity-RV decorrelation; (2) A two-planet model without any decorrelation; and (3) A two-planet model with activity-RV decorrelation. We compare these three models in Table \ref{tab:hd99492_results}.

\begin{table}[h]
\fontsize{6}{7.2}
\centering
\caption{Model Comparison for HD 99492 \label{tab:hd99492_results}}
\begin{tabular}{lrrr}
\hline
\hline
& \textbf{Model 1} 	& \textbf{Model 2} & \textbf{Model 3} \\
& \textbf{(Adopted)} & & \\
\hline
Number of planets 	& 1		& 2		& 2		\\
\Shk correction			& Full 	& None 	& Full 	\\
\hline
RMS residuals (\ms)		& 3.61 		& 3.69 		& 3.39 		\\
BIC						& 502.98 	& 524.53 	& 513.52 	\\
\hline
\hline
\end{tabular}
\end{table}

For the HD~99492 system, the lowest BIC was achieved for the model with only one planet, with the residual RV signal fully accounted for by stellar activity ($\dbic = -11$). The slope of the RV-\Shk correlation was found to be $c_1= 92 \pm 18$~\ms~\Shk$^{-1}$, so the semi-amplitude of the activity-induced RV signal is $\sim 5$~\ms, similar to that found for HAT-P-11. We also noted that when fitting two planets with activity decorrelation (model 3), an MCMC analysis found the RV semi-amplitude of the outer planet to be 1.8$^{+1.2}_{-2.0}$~\ms, not significantly different from zero. Finally, a Keplerian fit to the \Shk time series found a cycle whose period and phase are consistent with the fit to the RV time series within 1-$\sigma$, suggesting that the two signals are indeed correlated.

Here, we see that the same analysis as previously applied to HAT-P-11 now readily rejects the presence of an outer planet in the HD 99492 system and supports the integrity of our methodology.

\section{Posterior distribution of RV model parameters}
\label{appendix:posteriors}

\begin{figure*}[h!]
	\plotone{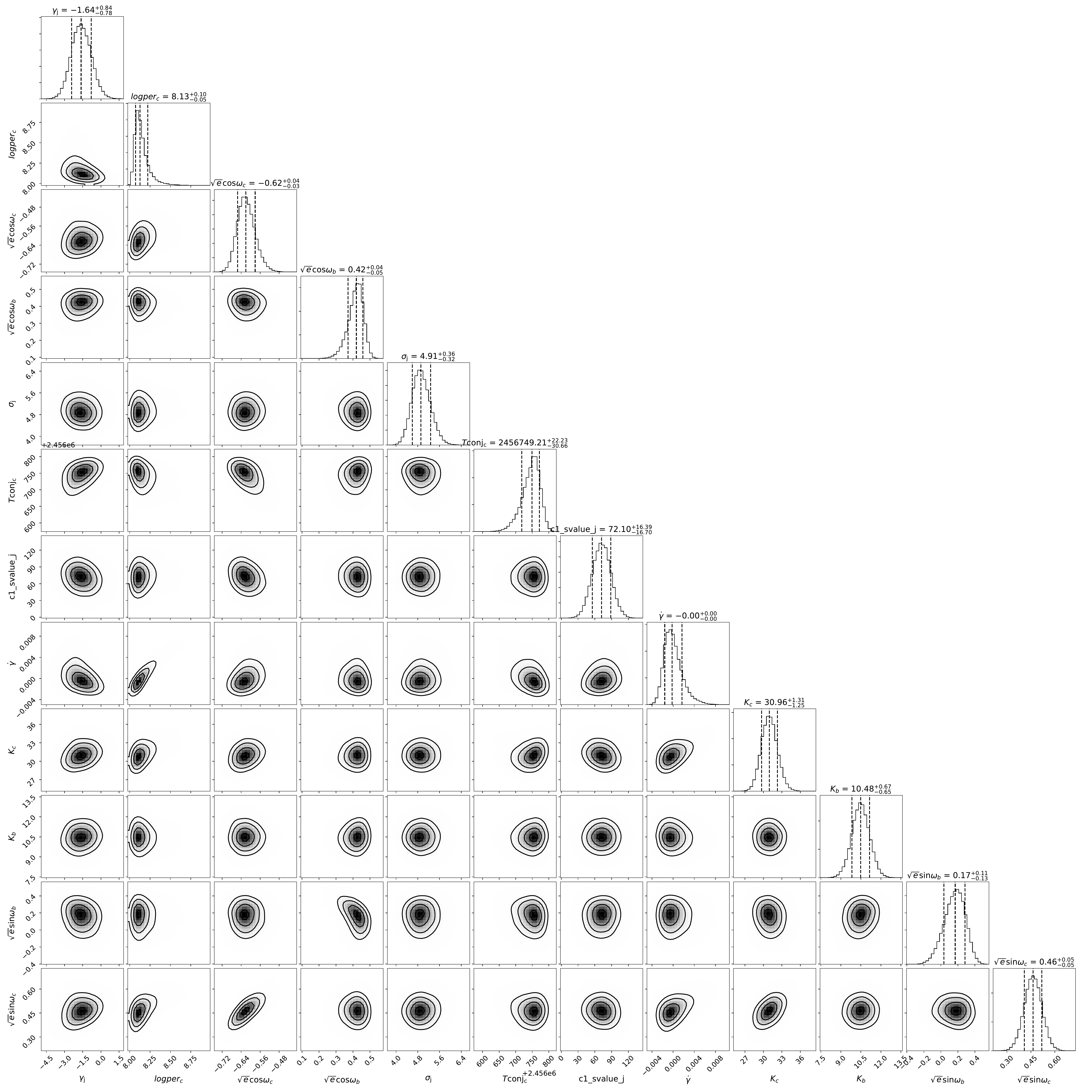}
	\caption{Corner plot showing posterior distributions for each model parameter. The first plot in each column shows the single-variable distribution, with the vertical dashed lines denoting the most probable value and 1$\sigma$ confidence bounds. The remaining plots show joint distributions between each pair of model parameters. \label{fig:corner}}
\end{figure*}

We provide in Figure \ref{fig:corner} the posterior distributions for each of the model parameters. To derive these distributions, we explored the likelihood surface with an MCMC analysis. 

\end{document}